\documentclass[a4paper]{jpconf}
 \usepackage{listings}             

\usepackage[T1]{fontenc}
\usepackage[latin9]{inputenc}
\usepackage{color}
\usepackage{amstext}
\usepackage{amssymb}
\usepackage{graphicx}
\usepackage{hyperref}
\usepackage{wasysym}
\usepackage{subfig}

\begin{document}
\lstset{language=Python}          
\title{Dynamic connectivity algorithms for Monte Carlo simulations of the random-cluster model}

\author{Eren Metin El\c{c}i and Martin Weigel}
\address{Applied Mathematics Research Centre, Coventry University, Coventry, CV1 5FB, United Kingdom and\\
  Institut f\"ur Physik, Johannes Gutenberg-Universit\"at Mainz, Staudinger Weg 7,
  D-55099 Mainz, Germany}
\ead{elcie@uni.coventry.ac.uk, Martin.Weigel@coventry.ac.uk}

\begin{abstract}
  We review Sweeny's algorithm for Monte Carlo simulations of the random cluster
  model. Straightforward implementations suffer from the problem of computational
  critical slowing down, where the computational effort per edge operation scales
  with a power of the system size.  By using a tailored dynamic connectivity
  algorithm we are able to perform all operations with a poly-logarithmic
  computational effort. This approach is shown to be efficient in keeping online
  connectivity information and is of use for a number of applications also beyond
  cluster-update simulations, for instance in monitoring droplet shape
  transitions. As the handling of the relevant data structures is non-trivial, we
  provide a Python module with a full implementation for future reference.
\end{abstract}

\section{Introduction}

The cluster-update algorithm introduced for simulations of the Potts model by
Swendsen and Wang in 1987 \cite{swendsen-wang:87a} has been a spectacular success,
reducing the effect of critical slowing down by many orders of magnitude for the
system sizes typically considered in computer simulation studies. A number of
generalizations, including an algorithm for continuous-spin systems and the
single-cluster variant \cite{wolff:89a} as well as more general frameworks for
cluster updates \cite{edwards:88a,kandel:91a}, have been suggested following the
initial work of Ref.~\cite{swendsen-wang:87a}. The single bond update introduced by
Sweeny \cite{sweeny:83} several years before Swendsen's and Wang's work is
considerably less well known. This is mostly due to difficulties in its efficient
implementation in a computer code, which is significantly more involved than for the
Swendsen-Wang algorithm. In deciding about switching the state of a given bond from
inactive to active or {\em vice versa\/}, one must know the consequences of the move
for the connectivity properties of the ensemble of clusters, i.e., whether two
previously disjoint clusters will become connected or an existing cluster is broken
up by the move or, instead, the number of clusters will stay unaffected. If
implemented naively, these connectivity queries require a number of steps which is
asymptotically close to proportional to the number of spins, such that the resulting
{\em computational critical slowing down\/} outweighs the benefit of the reduced
autocorrelation times of the updating scheme. Even though it was recently shown that
the decorrelation effect of the single-bond approach is asymptotically {\em
  stronger\/} than that of the Swendsen-Wang approach \cite{deng:07,elci:13}, this
strength can only be played once the computational critical slowing down is brought
under control. Here, we use a poly-logarithmic dynamic connectivity algorithm as
recently suggested in the computer science literature
\cite{henzinger:99,holm:01,iyer:01} to perform bond insertion and removal operations
as well as connectivity checks in run-times only logarithmically growing with system
size, thus removing the problem of algorithmic slowing down. As the mechanism as well
as the underlying data structures for these methods are not widely known in the
statistical physics community, we here use the opportunity to present a detailed
description of the approach. For the convenience of the reader, we also provide a
Python class implementing these codes, which can be used for simulations of the
random-cluster model or rather easily adapted to different problems where dynamic
connectivity information is required.

\section{\label{sec:model_sy} The random-cluster model and Sweeny's algorithm}

We consider the random-cluster model (RCM) \cite{grimmett:book} which
is a generalization of the bond percolation problem introducing a correlation between
sites and bonds. It is linked to the $q$-state Potts model through the
Fortuin-Kasteleyn transformation \cite{fortuin:72a,fortuin:72b,fortuin:72c},
generalizing the Potts model to arbitrary real $q>0$. Special cases include regular,
uncorrelated bond percolation ($q=1$) as well as the Ising model ($q=2$). To define
the RCM, consider a graph $\mathcal{G}\equiv(V,E)$ with vertex set $V$, ($\vert
V\vert\equiv N$), and edge set $E$, ($\vert E\vert\equiv M$). We associate an
occupation variable $\omega(e)\in\{0,1\}$ with every edge $e\in E$. We say that $e$
is \emph{active} if $\omega(e)=1$ and \emph{inactive} otherwise. The state space
$\Omega$ of the RCM corresponds to the space of all (spanning\footnote{A subgraph $\mathcal{A}\subseteq G$ is spanning if it contains all vertices of $G$.}) sub-graphs,
\begin{equation}
    \Omega\equiv \lbrace \mathcal{A}=(V,A) \vert A\subseteq E  \rbrace=\lbrace0,1\rbrace^{M}.
    \label{eq:statespace}
\end{equation}
A configuration is thus represented as
$\vec{\omega}\equiv[\omega(e_{1}),\omega(e_{2}),...,\omega(e_{M})]\in\Omega$ and
corresponds uniquely to a sub-graph
$\mathcal{A(\vec{\omega})}\equiv(V,A(\vec{\omega}))\subseteq\mathcal{G}$ with
\begin{equation}
    A(\vec{\omega})=\{e\in E\vert\omega(e)=1\}\subseteq E.
\label{eq:subgraph}
\end{equation}
The probability associated with a configuration $\vec{\omega}\in\Omega$
is given by the RCM probability density function (PDF)
\begin{eqnarray}
    \mu(p,q,\vec{\omega}) & \equiv &\frac{1}{Z(p,q)} \left[\prod_{e\in E}p^{\omega(e)}(1-p)^{1-\omega(e)}\right]q^{k(\vec{\omega})},
    \label{eq:rcm_pmf}\\
    Z(p,q) & \equiv & \sum_{\vec{\omega}\in\Omega} \left[\prod_{e\in E}p^{\omega(e)}(1-p)^{1-\omega(e)}\right]q^{k(\vec{\omega})}
    \label{eq:rcm_pmf_2},
\end{eqnarray}
where $k(\vec{\omega})$ is the number of connected components (clusters) and $Z(p,q)$
denotes the RCM partition function. More generally, as a function of the parameters
$p\in[0,1]$, the density of active edges, and $q\in(0,\infty)$, the cluster number
weight, these expressions define a family of PDFs.  It is worthwhile to mention a
number of limiting cases.  For $q\rightarrow 1$ Eq.~(\ref{eq:rcm_pmf}) factorizes and
corresponds to independent bond percolation with $Z(p) \rightarrow 1$. In the limit
of $q\rightarrow 0$ with fixed ratio $w=v/q$, on the other hand, it corresponds to
bond percolation with local probability $w/(1+w)$ and the condition of cycle-free
graphs. Taking $w\rightarrow \infty$ or in the limit of $q\rightarrow 0$ and
$v/q^\sigma$ constant for $0<\sigma<1$ we obtain the ensemble of uniform spanning
trees for connected $G$. Naturally, in the latter two limits every edge in a
configuration is a bridge.

Sweeny's algorithm \cite{sweeny:83} is a local bond updating algorithm
directly implementing the configurational weight (\ref{eq:rcm_pmf}). We first
consider its formulation for the limiting case $q\to 1$ of independent bond
percolation. For an update move, randomly choose an edge $e\in E$ with uniform
probability and propose a flip of its state from inactive to active or {\em vice
  versa\/}. Move acceptance can be implemented with any scheme satisfying detailed
balance, for instance the Metropolis acceptance ratio $\min{(v^{\Delta w},1)}$ where
$v\equiv p/(1-p)$ and $\Delta \omega = \pm 1$ for insertions and deletions of edges,
respectively. This dynamical process is described by the following master equation:
\begin{equation}
    \mathbb{P}(\vec{\omega},t+1)=(1-r(\vec{\omega}))\mathbb{P}(\vec{\omega},t)+\sum_{\vec{\omega}'\neq \vec{\omega} }W(\vec{\omega}'\rightarrow\vec{\omega})\mathbb{P}(\vec{\omega'},t),
\label{eq:master_eq}
\end{equation}
where  $r(\vec{\omega}) = \sum_{\vec{\omega}' \neq \vec{\omega}} W(\vec{\omega} \rightarrow
\vec{\omega'})$ ensures proper normalization of  $\mathbb{P}(\vec{\omega},t+1)$, given the normalization of $\mathbb{P}(\vec{\omega},t)$ . The Metropolis transition rates are then given by
\begin{eqnarray}
    W(\vec{\omega}  \rightarrow  \vec{\omega}')&\equiv&\frac{1}{M}\sum_{m=1}^{M}W_{m}(\vec{\omega}\rightarrow\vec{\omega}')
    \label{eq:glob_transr},
    \\ 
    W_{m}(\vec{\omega}\rightarrow\vec{\omega}') & \equiv & \min{(1,v^{\Delta w})} \prod_{l\neq m}\left[\delta_{\omega_{l},\omega_{l}'}\right]
    \label{eq:loc_transr}.
\end{eqnarray}
Eq.~(\ref{eq:glob_transr}) expresses the uniform random selection of an edge and the
corresponding edge dependent transition rate $W_{m}$ is defined in
Eq.~(\ref{eq:loc_transr}). The product of Kronecker deltas ensures the single-bond
update mechanism, i.e., that only one edge per step is changed. From here,
generalization to arbitrary $q\in(0,\infty)$ is straightforward, leading to a
modified transition rate
\begin{eqnarray}
  W_{m}(\vec{\omega} & \rightarrow & \vec{\omega}')\equiv\ \min(1,q^{\Delta k}v^{\Delta\omega})\prod_{l\neq m}\left[\delta_{\omega_{l},\omega_{l}'}\right].
  \label{eq:transr_rcm}
\end{eqnarray}
We note that this Metropolis update is more efficient than a heat-bath variant for
any value of $q$ apart from $q=1$, where both rules coincide. Clearly, for $q\ne 1$,
to compute the acceptance probability $W_{m}$ of a given trial move one must find
$\Delta k$, the change in connected components (clusters) induced by the move. This
quantity, equivalent to the question of whether the edge $e$ is a {\em bridge\/}, is
highly non-local. Determining it involves deciding whether there exists at least one
alternative path of active edges connecting the incident vertices $x$ and $y$ that
does not cross $e = (x,y)$.

\section{\label{sec:dc_algo}The connectivity problem}

The dynamic connectivity problem is the task of performing efficient connectivity
queries to decide whether two vertices $x$ and $y$ are in the same ($x\leftrightarrow
y$) or different ($x\nleftrightarrow y$) connected components for a dynamically
evolving graph, i.e., mixing connectivity queries with edge deletions and
insertions. For a static graph, such information can be acquired in asymptotically
constant time after a single decomposition, for instance using the Hoshen-Kopelman
algorithm \cite{hoshen:79}. Under a sequence of edge insertions (but no deletions),
it is still possible to perform all operations, insertions and connectivity queries,
in practically constant time using a so-called union-and-find (UF) data structure
combined with path-compression and tree-balance heuristics
\cite{cormen:09}. This fact has been used to implement a very efficient
algorithm for the (uncorrelated) percolation problem \cite{newman_fast_2001}. An
implementation of Sweeny's algorithm, however, requires insertions as well as
deletions to ensure balance. Hence, we need to be able to remove edges without the
need to rebuild the data structure from scratch.

\begin{table}[bt]
  \centering
  \caption{Asymptotic run-time scaling at criticality of the elementary operations of insertion or
    deletion of internal or external edges, respectively, using sequential
    breadth-first search (SBFS), interleaved BFS (IBFS), union-and-find (UF) or the
    fully dynamic connectivity algorithm (DC) as a function of the linear system size
    $L$.}
  \label{tab:scaling}
    \begin{tabular}{lllll}
      \multicolumn{1}{c}{operation} & \multicolumn{1}{c}{SBFS} &  \multicolumn{1}{c}{IBFS} &
      \multicolumn{1}{c}{UF} & \multicolumn{1}{c}{DC} \\  \hline
      internal insertion & $L^{d_F-x_2}$          & $L^{d_F-x_2}$ & const.         & $\log^2 L$ \\
      external insertion & $L^{\gamma/\nu}$& $L^{d_F-x_2}$ & const.         &$\log^2 L$\\
      internal deletion  & $L^{d_F-x_2}$          & $L^{d_F-x_2}$ & $L^{d_F-x_2}$   &$\log^2 L$\\
      external deletion  & $L^{\gamma/\nu}$& $L^{d_F-x_2}$ & $L^{\gamma/\nu}$ &$\log^2 L$\\ \hline
      dominant & $L^{\gamma/\nu}$ &  $L^{d_F-x_2}$ & $L^{\gamma/\nu}$ &$\log^2 L$\\
    \end{tabular}
\end{table}

This goal can be reached using a number of different techniques. Building on the
favorable behavior of the UF method under edge insertions and connectivity queries,
the data structure can be updated under the removal of an external edge (bridge) by
performing breadth-first searches (BFSs) through the components connected to the two
ends $x$ and $y$ of the edge $e=(x,y)$. Alternatively, one might try to do without
any underlying data structure, answering each connectivity query through a separate
graph search in breadth-first manner. In both cases, the process can be considerably
sped up by replacing the BFSs by {\em interleaved\/} traversals alternating between
vertices on the two sides of the initial edge and terminating the process as soon as
one of the two searches comes to an end \cite{weigel:10d,deng:10,elci:13}. As, at
criticality of the model, the sizes of the two cluster fragments in case of a bridge
bond turn out to be very uneven on average, this seemingly innocent trick leads to
dramatic run-time improvements \cite{elci:13}. The asymptotic run-time behavior of
insertion and deletion steps for internal and external edges and the algorithms based
on BFS or UF data structures is summarized in Table \ref{tab:scaling} for the case of
simulations on the square lattice of edge length $L$. We expect the same bounds with
the corresponding exponents to hold for general critical hypercubic lattices. Here,
$\gamma/\nu$ denotes the finite-size scaling exponent of the susceptibility and
$d_F-x_2$ is a geometric exponent related to the two-arm crossing behavior of
clusters \cite{deng:10}. We note that $d_F-x_2 < \gamma/\nu$ for $0 < q \le 4$ in two
dimensions. Asymptotically, it is the most expensive operation which dominates the
run-time of the algorithm and, as a consequence, it turns out that (for the square
lattice) a simple BFS with interleaving is more efficient than the approach based on
union-and-find, cf.\ Table \ref{tab:scaling}.

In any case, the implementations discussed so far feature a computational effort for
a sweep of bond updates that scales faster than linearly with the system size, thus
entailing some computational critical slowing down. It is found in
Ref.~\cite{elci:13} that for most choices of $q$, this effect appears to
asymptotically destroy any advantage of a faster decorrelation of configurations by
the Sweeny algorithm as compared to the Swendsen-Wang method. An alternative
technique based on more complicated data structures allows to perform any mix of edge
insertions, deletions and connectivity queries in \emph{poly-logarithmic} run-time
per operation \cite{henzinger:99,holm:01,iyer:01}. Poly-logarithmic here denotes
polynomials of powers of the logarithm of the independent variable, e.g., system size
$L$, of the form
\begin{equation}
f(L)=\sum_{k=0}^{K}\alpha_{k}\log^kL .
\label{eq:polylog}
\end{equation}
Here the base of the logarithm is not important and changes only the coefficients. In
the following all logarithms are with respect to base $2$. Given the observation of
generally faster decorrelation of configurations by the Sweeny algorithm in the sense
of smaller dynamical critical exponents \cite{deng:07,elci:13}, the use of such
(genuinely) dynamic connectivity algorithms (DC) allows for an asymptotically more
efficient simulation of the critical random-cluster model \cite{elci:13}. In the
following, we discuss the basic ideas and some details of the algorithm employed
here.

\subsection{Trees, Forests and Euler tours}

\begin{figure}[tb]
  \centering{}\subfloat[Level $0$.]{\label{spanning_graph_0}\includegraphics[width=0.36\columnwidth]{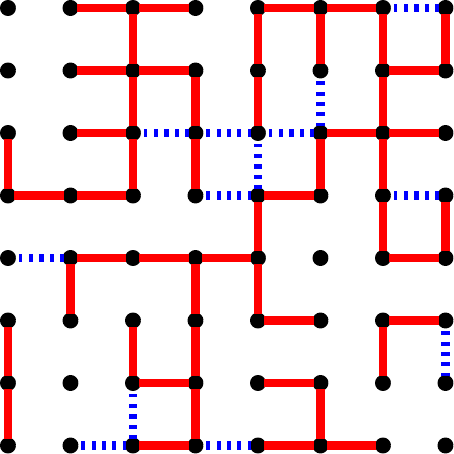}}\hspace{0.1\columnwidth}\subfloat[Level $1$.]{\label{spanning_graph_1}\includegraphics[width=0.36\columnwidth]{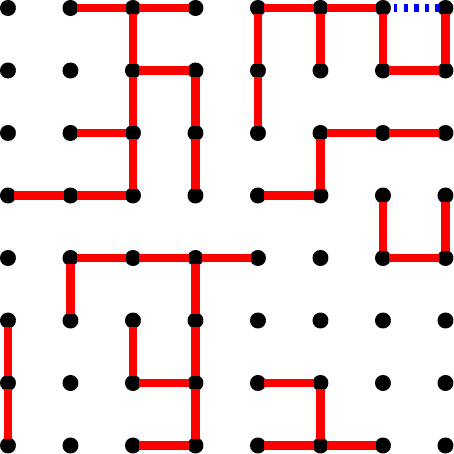}}
  
  \caption{A sub-graph $\mathcal{A}$ of a $8\times8$ square lattice. The solid (red)
    lines correspond to edges in one spanning forest $\mathcal{F}(\mathcal{A})$ and
    the dashed (blue) lines are additional edges not in the current spanning
    forest. Note that the graph has periodic boundary conditions and hence there are
    edges wrapping around the lattice horizontally and vertically. For the sake
    of clarity these are not shown.}
\end{figure}

The DC algorithm is based on the observation that for a given sub-graph $\mathcal{A}$
it is possible to construct a spanning forest $\mathcal{F}(\mathcal{A})$ which is defined by the
following properties:
\begin{itemize}
    \item $x\leftrightarrow y$ in $\mathcal{F}(\mathcal{A})$ if and only if $x\leftrightarrow y$ in $\mathcal{A}$
    \item there exists exactly one path for every pair $x$, $y$ with $x\leftrightarrow y$
\end{itemize}
In other words a spanning forest of a graph associates a spanning tree to every
component, an example is given in Fig.~\ref{spanning_graph_0} (solid lines only). One 
advantage of $\mathcal{F}(\mathcal{A})$ is that it has fewer edges than ${\cal A}$,
but represents the same connectivity information. For the sub-graph ${\cal A}$, the
distinction of {\em tree edges\/} $e\in\mathcal{F}(\mathcal{A})$ and {\em non-tree
  edges\/} $e\in\mathcal{A}\backslash\mathcal{F}(\mathcal{A})$ allows for a cheap
determination of $\Delta k$. For the case of deleting an edge in
$e\notin\mathcal{F}(A)$ we know that there is an alternative path connecting the
adjacent vertices, namely the path in $\mathcal{F}(\mathcal{A})$, so this edge was 
part of a cycle and we conclude $\Delta k=0$. If we insert an edge whose adjacent
vertices are already connected in $\mathcal{F}(\mathcal{A})$ then we come to the same
conclusion.

\begin{figure}[bt]
    \centering
    \raisebox{0cm}{
      \includegraphics[scale=.8]{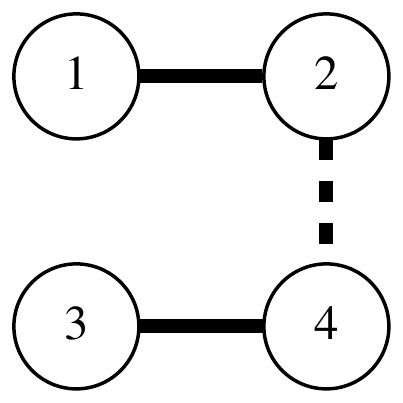}
    }
    \hspace*{0.5cm}
    \raisebox{1.3cm}{
      \scalebox{2}{\LARGE $\mapsto$}
    }
    \hspace*{0.5cm}
   \raisebox{1.1cm}{
      \includegraphics[scale=.8]{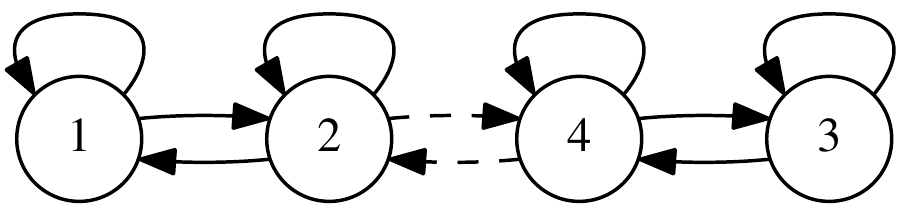}
    }
    \hspace*{0.5cm}
    \caption{ \label{circuit_mapping}Mapping of a component to a directed circuit
      with loops. The dashed line is an edge to be deleted. The dashed arrows will be
      removed to update the directed circuit.}
\end{figure}

If, on the other hand, we want to insert a tree edge, i.e., an edge with adjacent
vertices $x$, $y$ not yet connected, we observe that because $x$ and $y$ belong to
separate spanning trees before the insertion of $e$, the new spanning subgraph
obtained by linking $x$ and $y$ via $e$ is still a spanning tree. Hence the only
modification on the spanning forest for the insertion of a tree edge is the
amalgamation of two trees. This can be done in $\mathcal{O}(\log L)$ steps by using the following
idea of Ref.~\cite{tarjan_dynamic_1997} which also supports the deletion of
bridges. For a given component $\mathcal{C}$ in $\mathcal{A}$ we transform the
corresponding tree $\mathcal{T}(\mathcal{C})$ in $\mathcal{F}(\mathcal{A})$ into a
directed circuit by replacing every edge $(x,y)$ by two directed edges (arcs) $[x,y]$
and $[y,x]$ and every vertex by a loop $[x,x]$. Figure \ref{circuit_mapping}
illustrates how to translate edge insertions or deletions into/from the spanning
forest to modifications on the directed circuits. Deleting an edge from a tree splits
it into two trees. The directed circuit therefore splits into two circuits corresponding to the two trees. 
When inserting a tree edge, i.e., merging two trees,
we join the circuit by the two arcs, corresponding to $e$, at the vertices incident to $e$.

By storing the directed circuits for every component in so called Euler tour
sequences (ETS), \cite{henzinger:99}, all necessary manipulations on the
directed circuits can be done in $\mathcal{O}(\log L)$ operations if we store each
ETS in a separate balanced search tree such as, for instance, a red-black, AVL, or B
tree \cite{knuth_art_1998}.  Alternatively one can use self-adjusting binary search
trees, so called splay trees \cite{sleator_self-adjusting_1985}. In this case the
bound is amortized, i.e., averaged over the complete sequence of operations. Due to a somewhat
simpler implementation and the fact that a Monte Carlo simulation usually consists of
millions of operations in random order naturally leading to amortization, we
concentrated on the splay-tree approach. Based on the ETS representation we can
translate connectivity queries into checking the underlying search tree roots for
equivalence.  An in-depth discussion of the exact manipulations on the representing
ETS is beyond the scope of this article and we refer the interested reader to the
literature \cite{henzinger:99}. Here we restrict ourselves to
considering the deletion of a bridge as an example.  The initial graph is the one
illustrated in the left panel of Fig.~\ref{circuit_mapping}. One corresponding ETS
sequence of the linearized directed cycle in the right panel of
Fig.~\ref{circuit_mapping} is the following:
\begin{equation}
\mathcal{E}=[1,1]\rightarrow[1,2]\rightarrow[2,2]\rightarrow[2,4]\rightarrow[4,4]\rightarrow[4,3]\rightarrow[3,3]\rightarrow[3,4]\rightarrow[4,2]\rightarrow[2,1].
\label{eq:sequence}
\end{equation}
Suppose now that we delete edge $(2,4)$ from the tree (cf.\ the dashed line in
Fig.~\ref{circuit_mapping}). This will result in a split of the component
$\mathcal{C}\equiv\{1,2,3,4\}$ into two parts $\mathcal{C}_{1}\equiv\{1,2\}$ and
$\mathcal{C}_{2}\equiv\{3,4\}$. In this case the deletion of edge $(2,4)$ translates
into a cut of the original sequence $\mathcal{E}$ at the two arcs $[2,4]$ and
$[4,2]$. The sequence $\mathcal{E}_2$ of arcs between these two edges corresponds to
$\mathcal{C}_{2}$ and the concatenation of the remaining sequences without the two
arcs corresponding to $(2,4)$ results in $\mathcal{E}_1$ representing
$\mathcal{C}_{1}$:
\begin{eqnarray}
    \mathcal{E}_{1} & = & [1,1]\rightarrow[1,2]\rightarrow[2,2]\rightarrow[2,1], \label{eq:et_1}\\
    \mathcal{E}_{2} &=  & [4,4]\rightarrow[4,3]\rightarrow[3,3]\rightarrow[3,4]. \label{eq:et_2}
\end{eqnarray}
In summary, we see that by mapping every component in $\mathcal{A}$ to a tree in
$\mathcal{F}(\mathcal{A})$ and every such tree to a directed circuit which we store
in an ETS we are able to perform edge insertions/deletions into/from $\mathcal{F}(A)$
as well as connectivity queries with an amortised $\mathcal{O}(\log L)$ computational
effort.

\subsection{Edge hierarchy}

The remaining operation not implemented efficiently by the provisions discussed so
far is the deletion of edges from $\mathcal{F}(\mathcal{A})$ which are not bridges,
i.e., for which a replacement edge exists outside of the spanning forest. The DC
algorithm first executes the tree splitting as in the case of a bridge
deletion. Additionally, however, it checks for a reconnecting edge in the set of
non-tree edges. If such an edge is found, it concludes $\Delta k=0$ and merges the
two temporary trees as indicated above by using the located non-tree edge, which
hence now becomes a tree edge.  If, on the other hand, no re-connecting edge is
found, no additional work is necessary as the initially considered edge is a
bridge. To speed up the search for replacement edges, we limit it to the smaller of
the two parts $\mathcal{C}_{1}$ and $\mathcal{C}_{2}$ as all potential replacement
edges must be incident to both components. To allow for efficient searches for
non-tree edges incident to a given component using the ETS representation, the
search-tree data structures are augmented such that the loop arc for every vertex
stores an adjacency list of non-tree edges (vertices) incident to it. Further, every
node in the underlying search tree representing the ETS carries a flag indicating if
any non-tree edge is available in the sub-tree [which is a sub-tree of the search
tree and not of $\mathcal{F}(\mathcal{A})$]. This allows for a search of replacement
edges using the Euler tour and ensures that any non-tree edge can be accessed in
$\mathcal{O}(\log L)$ time.

It turns out that exploiting this observation is, in general, beneficial but not
sufficient to ensure the amortized time complexity bound indicated for the DC
algorithm in general. Suppose that the graph consists of a giant homogeneous
component with $M=\alpha N$ and $0 < \alpha \leq 1$ and the edge deletion results,
temporarily setting aside the question of possible replacement edges, in two trees
with $ \alpha_1 N$ and $(\alpha -\alpha_1) N$ incident edges, respectively, where
$0<\alpha_1 \leq \alpha/ 2$.  Then the computational effort caused by scanning all
possible non-tree edges is clearly $\mathcal{O}(N)$. In amortizing onto the
insertions performed to build up this component, every such non-tree edge carries a
weight of $\mathcal{O}(\log L)$.  If this case occurs sufficiently frequently, it
will be impossible to bound the amortized cost per operation.  This problem is
ultimately solved in the DC algorithm by the introduction of an edge hierarchy. The
intuitive idea is to use the expense of a replacement edge search following a
deletion to reduce the cost of future operations on this edge.  This is done in such
a way as to separate dense from sparse clusters and more central edges from those in
the periphery of clusters. By amortizing the cost of non-tree edge scans and level
increases over edge insertions it follows that one can reduce the run-time for graph
manipulations to an amortized $\mathcal{O}(\log^2 L)$ and $\mathcal{O}(\log L)$ for
connectivity queries \cite{holm:01}. Each time an incident non-tree edge is checked
and found unsuitable for reconnecting the previously split cluster we promote it to
be in a higher level. If we do this many times for a dense component we will be able
to find incident non-tree edges very quickly in a higher level. These ideas are
achieved in the DC algorithm by associating a level function to each $e\in E$,
\begin{equation}
  0\leq\ell(e)\leq\ell_{\max}\equiv\lfloor\log N \rfloor.
  \label{eq:levels}
\end{equation}
Based on this level function, one then constructs sub-graphs
$\mathcal{A}_{i}\subseteq \mathcal{A}$ with the property
\begin{equation}
  \mathcal{A}_{i}\equiv\lbrace e\in\mathcal{A}\vert l(e)\ge i\rbrace.
\end{equation}
This induces a hierarchy of sub-graphs:
\begin{equation}
  \mathcal{A}_{\ell_{\max}}\subseteq\cdots\subseteq\mathcal{A}_{1}\subseteq\mathcal{A}_{0}\equiv\mathcal{A}.
  \label{eq:graph_hierarchy}
\end{equation}
As described above, for every sub-graph we construct a spanning forest
$\mathcal{F}_{i}\equiv\mathcal{F}(\mathcal{A}_{i})$. Clearly the same hierarchy holds
for the family of spanning forests. In other words the edges in level $i-1$ connect
components/trees of level $i$. If an edge has to be inserted into $\mathcal{A}$, then
it is associated to a level $l(e)=0$ and hence it is in $\mathcal{A}_{0}$. To achieve
an efficient search for replacement edges, the algorithm adapts the level of edges
after deletions of tree edges in a way which preserves the following two invariants
\cite{holm:01}:
\begin{enumerate}
\item The maximal number of vertices in a component in level $i$ is
  $\lfloor N/2^{i}\rfloor$.
\item Any possible replacement edge for a previously deleted edge $e$ with
level $l$ has level $\leq l$.
\end{enumerate}
Trivially, both invariants are fulfilled when all edges have level 0. We now have to
specify how exactly the idea of keeping important edges at low levels and unimportant
ones at higher levels is implemented. To do this, suppose we deleted an edge from
$\mathcal{F}_{i}$, i.e., at level $i$, and temporarily have
$\mathcal{T}\rightarrow\mathcal{T}_{1}+\mathcal{T}_{2}$ where (say) $\mathcal{T}_{1}$
is the smaller of the two, i.e., it has less vertices. Because of invariant (i) it
follows that we are allowed to move the tree $\mathcal{T}_{1}$ (which is now at most
half the size of $\mathcal{T}$) to level $i+1$ by increasing the level of all tree
edges of $\mathcal{T}_{1}$ by one.  After that we start to search for a replacement
edge in the set of non-tree edges stored in the ETS of $\mathcal{T}_{1}$ in level $i$
where it also remains because of the fact that
$\mathcal{F}_{i+1}\subseteq\mathcal{F}_{i}$. For every scanned non-tree edge we have
two options:
\begin{itemize}
\item It does not reconnect $\mathcal{T}_{1}$ and $\mathcal{T}_{2}$ and has therefore
  both ends incident to $\mathcal{T}_{1}$. In this case, we increase the level of
  this edge $i\rightarrow i+1$. This implements the idea of moving unimportant edges
  in ``dense'' components to higher levels.
\item It does reconnect and hence we re-insert it at level $0$.
\end{itemize}
If we have not found a replacement edge at level $i$ we continue at level $i-1$.
The search terminates after unsuccessfully completing the search at level $0$ or when
a replacement edge was found.  In the first case it follows
$\mathcal{C}\rightarrow\mathcal{C}_{1}+\mathcal{C}_{2}$ whereas in the second case
$\mathcal{C}$ remains unchanged.

Implementing this replacement-edge search following any tree-edge deletion introduces
an upward flow of edges in the hierarchy of graphs and the level of an edge in the
current graph never decreases. Focusing on a single edge, we see that it is
sequentially moved into levels of smaller cluster size and hence the cost of future
operations on this edge is reduced.  Taking this into account it follows that the
insertion of an edge has a cost of $\mathcal{O}(\log L)$ for inserting at level $0$
plus it also ``carries'' the cost of all possible $\mathcal{O}(\log L)$ level
increases with cost of $\mathcal{O}(\log L)$ each resulting in $\mathcal{O}(\log^2
L)$ amortized per insertion. Deletions on the other hand imply a split of cost
$\mathcal{O}(\log L)$ in $\mathcal{O}(\log L)$ levels.  In case of an existing
replacement edge another contribution of $\mathcal{O}(\log L)$ caused by an insertion
at level 0 is added. The contribution of moving tree edges to higher levels and
searching for replacement edges (moving non-tree edges up) is already paid for by the
sequence of previous insertions (amortization). The only missing contribution is the
$\mathcal{O}(\log L)$ effort for obtaining the next replacement edge in an ETS. In total,
deletions hence have an amortized computational cost of $\mathcal{O}(\log^2 L)$.

\subsection{\label{sec:mem}Performance and optimizations}

\begin{figure}[tb]
  \centering{}\includegraphics[width=0.6\columnwidth]{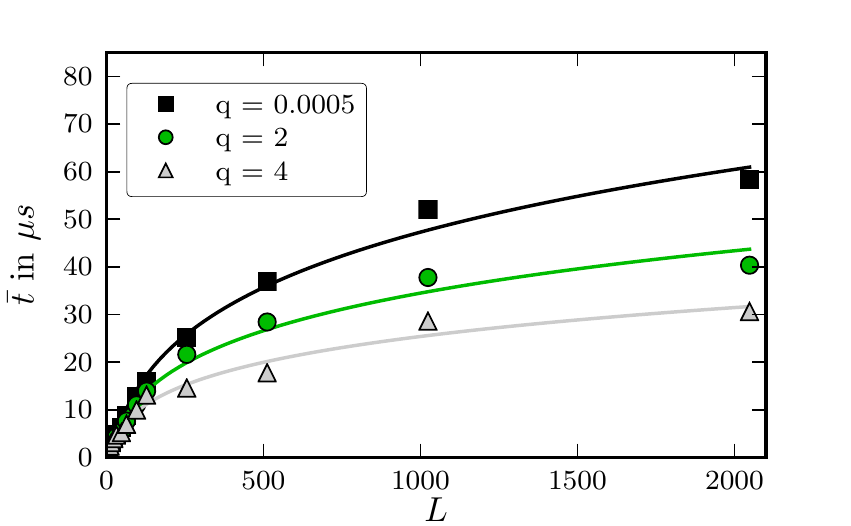}
  \caption{\label{fig:fig_dc_run_times}
    Average run-time for Sweeny's algorithm for the RCM for different values of $q$ at the critical
    point $v_c=\sqrt{q}$ on the 2D square lattice with periodic boundary conditions. }
\end{figure}

We tested the performance of the current DC implementation in the context of Sweeny's
algorithm in comparison to the simpler approaches based on breadth-first search and
union-and-find strategies. While the algorithm discussed here allows all operations
to be performed in poly-logarithmic time, due to the complicated data structures the
constants are relatively large. Our results show consistency with the
poly-logarithmic run-time bounds derived. It appears, however, that very large system
sizes are required to clearly see the superior asymptotic performance of the DC
algorithm as compared to the BFS and UF implementations. For details see the more
elaborate discussion in Ref.~\cite{elci:13}. As an example,
Fig.~\ref{fig:fig_dc_run_times} shows the average run-time per edge operation as a
function of the system size for three different choices of parameters.

Apart from run-time considerations, the implementation has a rather significant space
complexity. Since we maintain $\mathcal{O}(\log L)$ overlapping forests over the
$L^2$ vertices, the space complexity is $\mathcal{O}(L^2 \log L)$. A heuristic
suggested in Ref.~\cite{iyer:01} to decrease memory consumption is a
truncation of higher edge levels as these are, for the inputs or graphs considered in
our application, sparsely populated. We checked the impact on our implementation by
comparing run-times and memory consumptions for a truncation $\ell_{\max} \rightarrow
\ell_{\max}/2$.  We did not see any significant change in the run-time. On the other
hand we observed a reduction of almost a factor of two in the memory
consumption. This conforms to our observation that during the course of a simulation
almost no edges reached levels beyond $\approx 10$ for system sizes $L\lesssim 1024$
where the actual maximal level according to Eq.~(\ref{eq:levels}) is $\ell_{\max} =
20$.

Likewise, a number of further optimizations or heuristics are conceivable to improve
the typical run-time behavior. This includes a sampling of nearby edges when looking
for a replacement edge before actually descending into the edge level hierarchy
\cite{iyer:01}. A number of such heuristics and experimental
comparisons of fully and partially dynamics connectivity algorithms has been
discussed in the recent literature, see
Refs.~\cite{zaroliagis:02,demetrescu:06,demetrescu:10}. A full exploration of these
possibilities towards an optimal implementation of the DC class of algorithms for the
purpose of the Sweeny update is beyond the scope of the current article and forms a
promising direction for future extensions of the present work.

\section{\label{sec:python}Sweeny Python class}

We provide a Python class \cite{sweeny_github} encompassing four different
implementations of Sweeny's algorithm based on:
\begin{itemize}
\item sequential breadth-first searches (SBFS)
\item interleaved breadth-first searches (IBFS)
\item union-and-find with interleaved breadth-first searches (UF)
\item poly-logarithmic dynamic connectivities as discussed
  here (DC)
\end{itemize}
The package is built on top of a C library and it is therefore possible to use the
library in a stand-alone compiled binary. The necessary source code is also
provided. For more details see the related project documentation
\cite{sweeny_github}.  The source code is published under the MIT license
\cite{license}.  Here we give a basic usage example, which simulates the RCM with
$q=2$ (the Ising model) at $v_c=\sqrt{2}$, using an equilibration time of $1000$
sweeps, a simulation length of $10000$ sweeps, and random number seed $1234567$ using
the DC implementation:
\begin{lstlisting}[frame=single,caption=Example usage of Sweeny class.]
from sweeny import Sweeny
sy = Sweeny(q=2.,l=64,beta=np.log(1. + np.sqrt(2.)),coupl=1.,
cutoff=1000,tslength=10000,rngseed=1234567,impl='dc')
sy.simulate()
\end{lstlisting}
In order to extract an estimate, say, of the Binder cumulant  $R=\langle
\mathcal{S}_4 \rangle/\langle \mathcal{S}_2^2\rangle$ we need to retrieve
the time series for $\mathcal{S}_4$ and $\mathcal{S}_2$,
\begin{lstlisting}[frame=single,caption=Retrieving time series.]
sec_cs_moment= sy.ts_sec_cs_moment
four_cs_moment = sy.ts_four_cs_moment
sec_cs_moment *= sec_cs_moment
binder_cummulant = four_cs_moment.mean()/sec_cs_moment.mean()
\end{lstlisting}
Once an instance of the Sweeny class is created, it is easy to switch the algorithm
and parameters as follows:
\begin{lstlisting}[frame=single,caption=Switching algorithm and parameters.]
sy.init_sim(q=1.3,l=64,beta=np.log(1.+np.sqrt(1.3.)),coupl=1.,
cutoff=5000,tslength=50000,rngseed=7434,impl='ibfs')
\end{lstlisting}

\section{\label{sec:conc}Conclusions}

We have shown how to implement Sweeny's algorithm using a poly-logarithmic dynamic
connectivity method and we described the related algorithmic aspects in some
detail. We hope that the availability of the source code and detailed explanations
help to bridge the gap between the computer science literature on the topic of
dynamic connectivity problems and the physics literature related to MC simulations of
the RCM, specifically in the regime $q<1$.

The availability of an efficient dynamic connectivity algorithm opens up a number of
opportunities for further research. This includes studies of the tricritical value
$q_c(d)$ where the phase transition of the random-cluster model becomes discontinuous
for dimensions $d > 2$ \cite{hartmann:05,weigel:10d} as well as the nature of the
ferromagnetic-paramagnetic transition for $q\rightarrow 0$ and $d>2$
\cite{Deng2007Ferromagnetic}.

\ack
E.M.E. would like to thank P.\ Mac Carron for carefully reading the manuscript. 

\section*{References}


\begin{thebibliography}{10}
\expandafter\ifx\csname url\endcsname\relax
  \def\url#1{{\tt #1}}\fi
\expandafter\ifx\csname urlprefix\endcsname\relax\def\urlprefix{URL }\fi
\providecommand{\eprint}[2][]{\url{#2}}

\bibitem{swendsen-wang:87a}
Swendsen R~H and Wang J~S 1987 {\em Phys. Rev. Lett.\/} {\bf 58} 86--88

\bibitem{wolff:89a}
Wolff U 1989 {\em Phys. Rev. Lett.\/} {\bf 62} 361--364

\bibitem{edwards:88a}
Edwards R~G and Sokal A~D 1988 {\em Phys. Rev. D\/} {\bf 38} 2009

\bibitem{kandel:91a}
Kandel D and Domany E 1991 {\em Phys. Rev. B\/} {\bf 43} 8539--8548

\bibitem{sweeny:83}
Sweeny M 1983 {\em Phys. Rev. B\/} {\bf 27} 4445

\bibitem{deng:07}
Deng Y, Garoni T~M and Sokal A~D 2007 {\em Phys. Rev. Lett.\/} {\bf 98} 230602

\bibitem{elci:13}
El\c{c}i E~M and Weigel M 2013 {\em Phys. Rev. E\/} {\bf 88} 033303

\bibitem{henzinger:99}
Henzinger M~R and King V 1999 {\em J. ACM\/} {\bf 46} 502--516

\bibitem{holm:01}
Holm J, de~Lichtenberg K and Thorup M 2001 {\em J. ACM\/} {\bf 48} 723--760

\bibitem{iyer:01}
Iyer R, Karger D, Rahul H and Thorup M 2001 {\em J. Exp. Algorithmics\/} {\bf
  6} 4

\bibitem{grimmett:book}
Grimmett G 2006 {\em The random-cluster model\/} (Berlin: Springer)

\bibitem{fortuin:72a}
Fortuin C~M and Kasteleyn P~W 1972 {\em Physica\/} {\bf 57} 536--564

\bibitem{fortuin:72b}
Fortuin C~M 1972 {\em Physica\/} {\bf 58} 393--418

\bibitem{fortuin:72c}
Fortuin C~M 1972 {\em Physica\/} {\bf 59} 545--570

\bibitem{hoshen:79}
Hoshen J and Kopelman R 1976 {\em Phys. Rev. B\/} {\bf 14} 3438--3445

\bibitem{cormen:09}
Cormen T~H, Leiserson C~E, Rivest R~L and Stein C 2009 {\em Introduction to
  Algorithms\/} 3rd ed (Cambridge, MA: MIT Press)

\bibitem{newman_fast_2001}
Newman M~E~J and Ziff R~M 2001 {\em Phys. Rev. E\/} {\bf 64} 016706

\bibitem{weigel:10d}
Weigel M 2010 {\em Physics Procedia\/} {\bf 3} 1499--1513

\bibitem{deng:10}
Deng Y, Zhang W, Garoni T~M, Sokal A~D and Sportiello A 2010 {\em Phys. Rev.
  E\/} {\bf 81} 020102

\bibitem{tarjan_dynamic_1997}
Tarjan R~E 1997 {\em Math. Program.\/} {\bf 78}

\bibitem{knuth_art_1998}
Knuth D~E 1998 {\em {The Art of Computer Programming: Sorting and Searching v.
  3: The Classic Work Newly Updated and Revised}\/} 2nd ed (Addison Wesley)

\bibitem{sleator_self-adjusting_1985}
Sleator D~D and Tarjan R~E 1985 {\em J. {ACM}\/} {\bf 32} 652--686

\bibitem{zaroliagis:02}
Zaroliagis C~D 2002 {\em Experimental Algorithmics\/} ({\em Lecture Notes in
  Computer Science\/} vol 2547) ed Fleischer R, Moret B and Schmidt E (Springer
  Berlin Heidelberg) pp 229--278

\bibitem{demetrescu:06}
Demetrescu C and Italiano G~F 2006 {\em ACM Trans. Algorithms\/} {\bf 2}
  578--601

\bibitem{demetrescu:10}
Demetrescu C, Eppstein D, Galil Z and Italiano G~F 2010 ed Atallah M~J and
  Blanton M (Chapman \& Hall/CRC), chapter {\em Dynamic graph algorithms\/}, p~9

\bibitem{sweeny_github}
El\c{c}i E~M {Sweeny Python module at Github},
  \url{https://github.com/ernmeel/sweeny}

\bibitem{license}
 {MIT License}, \url{http://opensource.org/licenses/MIT}

\bibitem{hartmann:05}
Hartmann A~K 2005 {\em Phys. Rev. Lett.\/} {\bf 94} 050601

\bibitem{Deng2007Ferromagnetic}
Deng Y, Garoni T~M and Sokal A~D 2007 {\em Phys. Rev. Lett.\/} {\bf 98}
  030602

\end{thebibliography}

\providecommand{\newblock}{}

\end{document}